\documentclass[aps,pre,reprint,floatfix,superscriptaddress]{revtex4-1}
\usepackage{graphicx}
\graphicspath{{./figs/}}
\usepackage{amsmath}
\usepackage{amssymb}
\usepackage{float}

\usepackage{amsthm}
\newtheorem{theorem}{Theorem}
\newtheorem{definition}{Definition}

\usepackage{xcolor}
\usepackage{framed}

\definecolor{ao(english)}{rgb}{0.0, 0.5, 0.0}
\colorlet{shadecolor}{blue!3}

\begin{document}
\title{Manifold Learning for Organizing Unstructured Sets of Process Observations}
%
\author{Felix Dietrich}
\affiliation{Department of Chemical and Biomolecular Engineering, 
and Department of Applied Mathematics and Statistics, 
Johns Hopkins University, Baltimore, MD 21218, USA}
\author{Mahdi Kooshkbaghi}
\affiliation{Program in Applied and Computational Mathematics,
Princeton University, Princeton, NJ 08544, USA}
\author{Erik M.~Bollt}
\affiliation{Department of Mathematics, and Department of Electrical and Computer
  Engineering, Clarkson Center for Complex Systems Science, Clarkson University, Potsdam, NY 13699-5815, USA}
\author{Ioannis G.~Kevrekidis}
\email{yannisk@jhu.edu}
\affiliation{Department of Chemical and Biomolecular Engineering, 
and Department of Applied Mathematics and Statistics, 
Johns Hopkins University, Baltimore, MD 21218, USA}
\begin{abstract}
Data mining is routinely used to organize ensembles of short temporal observations so as to reconstruct useful, low-dimensional realizations of an underlying dynamical system. 
In this paper, we use manifold learning to organize unstructured ensembles of observations (``trials'') of a system's response surface.
We have no control over where every trial starts; and during each trial 
operating conditions are varied by turning ``agnostic'' knobs, which change
system parameters in a systematic but unknown way.
As one (or more) knobs ``turn''
we record (possibly partial) observations of the system response.
%
We demonstrate how such partial and disorganized observation ensembles can be integrated into coherent 
response surfaces whose dimension and parametrization can be systematically 
recovered in a data-driven fashion.
The approach can be justified through the Whitney and Takens embedding theorems, allowing reconstruction of manifolds/attractors through different types of observations.
We demonstrate our approach by organizing unstructured observations of response surfaces, including the reconstruction of a cusp bifurcation surface for Hydrogen combustion in a Continuous Stirred Tank Reactor.
%
%
Finally, we demonstrate how this observation-based reconstruction naturally leads to informative transport maps between input parameter space and output/state variable spaces.
\end{abstract}
\maketitle
\section{Introduction}
When an accurate mathematical model of a dynamical system is available, 
one can systematically observe the dependence of its response (its long-term dynamics, for example its steady states) on its parameters by computing the system bifurcation diagram/response surface through established numerical continuation and bifurcation packages like AUTO~\cite{doedel1981auto, doedel2007auto} or MATCONT~\cite{dhooge-2003}.
One starts from a well defined initial point on the response surface,
(e.g. a steady state at a particular set of parameter settings) and then builds the surface by systematically moving on it.
This exploration typically involves varying one parameter at a time (i.e. following one-dimensional curves on this surface, performing ``one-parameter continuation"). It is also possible to explore the response surface through simplicial continuation, systematically varying two (and possibly even more) parameters at a time. At every new point of a simplex the algorithms return the steady state values of each and every model variable.

In contrast to this ``complete control, full knowledge" scenario, we want to explore a scenario closer to what an experimentalist might observe when exploring a new, unknown or only partially understood system.
We may not be able to measure all components of the system state; during each ``trial" (each sequence of experiments) we can vary experimental ``knobs" that systematically change conditions, but in a way unknown to us; and finally, the location of the starting point of each trial in parameter and state space may also be unknown (e.g. set by uncontrolled environmental conditions). 

In this paper we will show that data mining the (partial) observations of the system response in such an uncontrolled (unstructured, agnostic) setting through manifold learning techniques like Diffusion Maps~\cite{Coifman2005}, can facilitate the construction of a meaningful realization of the correct response surface. 
This can subsequently be exploited to classify, analyze and even predict the system response to variations of our experimental ``knobs".

\begin{figure}[!htp]
\centering
\includegraphics[width=0.5\textwidth]{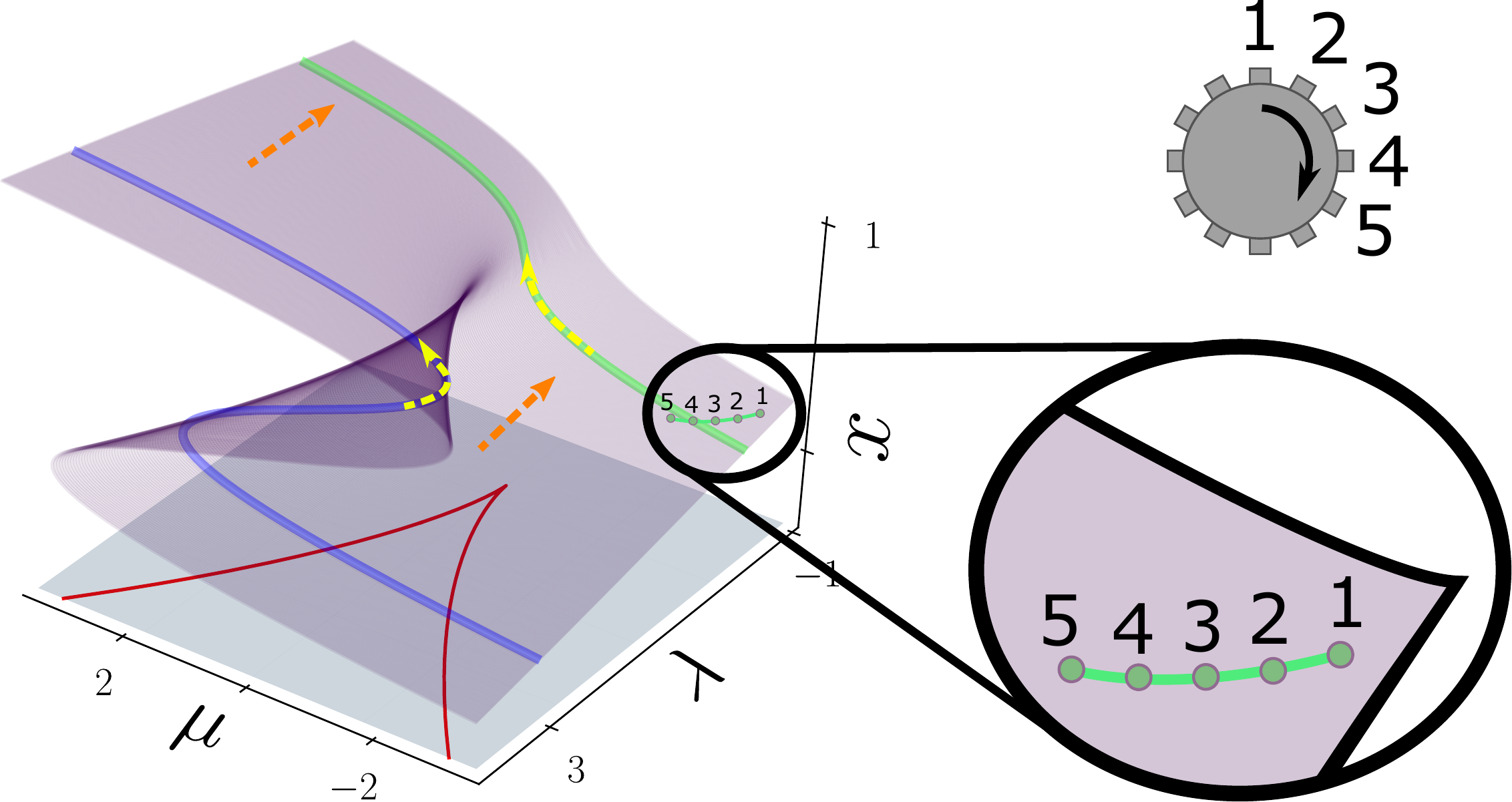}
\caption{
The cusp surface
embedded in parameters and state space 
$\mathbb{R}^2\times\mathbb{R}$.
The blue and green curves are observed at two constant  $\lambda$ values.
Yellow and orange arrows indicate short observation segments for each of the two 
typical one-parameter continuation directions along the surface.
For $\mu$ and $\lambda$ values located 
between the two red ``fold'' lines, the system exhibits
hysteresis.
The inset, with the green curve and the ``knob" illustrates our scenario in which only conditions along short curves on the surface can be visited in a systematic but agnostic manner.}
\label{fig:x_mu_lambda_2d}
\end{figure}

An instructive caricature of the procedure is illustrated in Fig.~\ref{fig:x_mu_lambda_2d}, in the form of the well-known cusp surface $0 = \mu + \lambda x - x^3$ 
where  a single state variable $x$ is depicted in relation to the two parameters $(\lambda,\mu)$.
If information about $(\lambda,\mu,x)$ in Fig.~\ref{fig:x_mu_lambda_2d} at every point (or a fine enough grid of points) on the surface is available, it is straightforward to analyze and visualize the entire two-dimensional surface in three dimensions.
A typical numerical computation would fix the value of all but one of the parameters. Here, we fix $\lambda$ along, say, the blue or the green curve, then continue the solution in the parameter $\mu$ along (yellow) one-parameter segments on these curves. 
In this context, direct observation of the surface 
would involve recording of $x$, $\mu$ and $\lambda$ triplets 
throughout the surface. 
The surface is the graph of the function 
$\mu=\mu(x, \lambda):=x^3-\lambda x$, and manifold learning techniques such as Diffusion Maps can easily parametrize it, as we describe below.

\begin{figure*}[!htp]
\centering
\includegraphics[width=0.95\textwidth]{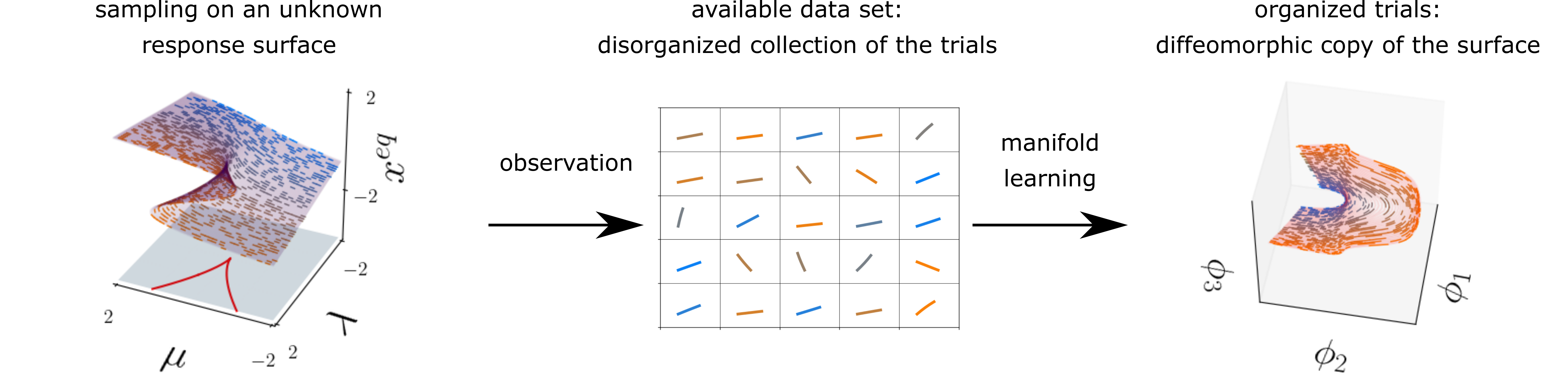}
\caption{
Observations of short trials from an unknown response surface (e.g., $x$ coordinates of the five points along each of the short lines, left) are available in a data set with no information about the global position of the lines (they are randomly shuffled, center). Manifold learning will organize the data set (right) into a diffeomorphic copy of the original surface. The paper explains and demonstrates that this approach can be employed for many different types of observations, as long as sufficiently rich local observation process histories are available.}
\label{fig:disorganized_data_manifold_learning}
\end{figure*}

In our more agnostic version of the exploration scenario, an experimenter cannot measure the parameter settings  $(\lambda,\mu)$ of the system, but is able to affect them through a single ``knob" that changes them, and through them changes the location of the state $x$ (Fig.~\ref{fig:x_mu_lambda_2d}).
Assume that the initial position of each trial,  (marked as ``1'') in the figure inset, is determined randomly by the environment, also without the knowledge of the experimenter. 
After the trial is initialized, for each shown angle $1-5$ of the knob, a corresponding point on the surface is visited, and the state value $x$ is recorded along the green curve segment depicted. For simplicity, one can consider, as we do here, that turning the knob at a constant rate moves the point visited at constant speed along the response surface; note that the approach does not rely on this particular assumption.
Repeating this tabulation with a large number of experiments from randomly distributed initial trials will result in a collection of recordings of five consecutive values of $x$ (see Fig.~(\ref{fig:disorganized_data_manifold_learning}), center)
without knowing the corresponding $\lambda, \mu$ values.

First, we demonstrate how such partial and disorganized observations of a response surface can be integrated in a coherent surface whose topology (the right ordering of the trials) and parametrization can be systematically recovered in a data-driven fashion.
Second, we extend the approach to different types of bifurcation observations (not just one-parameter continuation),  and demonstrate it in a more applied scenario with a Continuous Stirred Tank Reaction (CSTR) combustion problem. 
%
Finally, we demonstrate how this observation-based reconstruction naturally leads to the construction of transport maps between the input (parameter space) and the output (state variable space) of the system or model.

Figure~\ref{fig:disorganized_data_manifold_learning} illustrates the ensemble of {\em several short, disorganized, possibly partial} observation sequences one might obtain from a set of trials that densely sample the surface.
Reconstructing the entire bifurcation surface from such a set connects with ongoing research in data driven identification of dynamical systems from time-series, e.g.~
\cite{Sauer-1994,yair-2017, brunton-2016e}, where reconstructing useful phase space realizations and even dynamics from partial observations of time series has a long history \cite{moore-1981,takens1981detecting}.
%
%

%

\section{Reconstruction from process history}
We first consider the case that $\lambda$ is unknown but fixed (here, the blue curve in Figure~\ref{fig:x_mu_lambda_2d}) and our observation mode leads to a 
large, disorganized collection of points in $\mathbb{R}^m$, here $m=5$ values of $x$
along short segments of the blue curve. 
The parameters $\lambda$ and $\mu$ are not observed.
The central question becomes: To what extent can we reconstruct the blue curve from these trial records?
This is where the embedding theorems by Whitney and Takens (see Appendix A) become relevant: 
we consider the values of $x$ along the curve as analogous to time-delayed measurements along a temporal trajectory 
(where the angle of our parameter knob plays here the role of time);
the blue curve is one-dimensional, so following Whitney's theorem
five observations are sufficient ($m > 2d$, with $d=1$ the intrinsic 
and $m$ the embedding dimension)  to reconstruct an embedding of it, 
{\em even though we have no direct observation of $\mu$}; $\mu$ is inferred
only implicitly.

The number of delays necessary for an embedding in Takens' theorem depends on the dimension of the manifold, which might not be known beforehand. 
Keeping more delays results in an embedding in a Euclidean space that is higher-dimensional than necessary. To reduce this dimension,
we use a non-linear manifold learning technique, Diffusion Maps (DMAP)~\cite{Coifman2005}.
%
Given $N$ data points $\mathcal{D}=\{y_1, ..., y_N\}$ in ambient 
Euclidean space $\mathbb{E} = \mathbb{R}^{m}$ close to a smooth manifold $\mathcal{M}$, we construct a graph between the points, where connectivity is based on a cutoff Gaussian kernel similarity measure w.r.t. the Euclidean distance in the ambient space $\mathbb{E}$:
For a given scale parameter $\epsilon>0$, the similarity between two distinct points $y_i$ and $y_j$ in $\mathbb{E}$
is defined through 
$K_{ij}=k(y_i,y_j) = \chi_{[0,\delta]}(r)\rm{exp}\left(-r^2/\epsilon \right)$, where $r:=d(y_i,y_j)$ and $\chi_{[0,\delta]}$ is the indicator function on $[0,\delta]$, $\delta>0$ such that the kernel has compact support. Appropriate choices of the parameters $\epsilon$ and $\delta$ depend on the data~\cite{Coifman2005,berry-2015}.
The DMAP algorithm is based on the convergence of the
normalized graph Laplacian on the data to the Laplace--Beltrami operator on the manifold.
If the data points $\mathcal{D}$ are not sampled uniformly in $\mathcal{M}$, the matrix $K$ has to be normalized by an estimation of the density, $P_{ij}=\sum_{i=1}^{N} K_{ij}$, $\tilde{K}=P^{-\alpha}KP^{-\alpha}$ where $\alpha=0$ (no normalization, \cite{belkin-2003}) can be used in the case of uniform sampling, and $\alpha=1$ otherwise \cite{Coifman2005}.
The kernel matrix $\tilde{K}$ is normalized by the diagonal matrix $D \in \mathbb{R}^{(N \times N)}$, where $D_{ii}=\Sigma_{j=1}^{N} \tilde{K}_{ij}$ for $i=1,\cdots,N$.
The non-linear parametrization (embedding) of the manifold is then given by a certain number $L$ of eigenvectors of $A=D^{-1}\tilde{K} \in \mathbb{R}^{N \times N}$, scaled by their respective eigenvalue (and avoiding harmonics of previous eigenvectors \cite{dsilva-2018}).
The new embedding dimension $L$ may be much smaller than the previous ambient space dimension $m$, in which case DMAP achieves dimensionality reduction.

\subsection{Reconstructing a curve}
\begin{figure}[!htp]
\centering
\begin{tabular}{cc}
\includegraphics[width=0.23\textwidth]{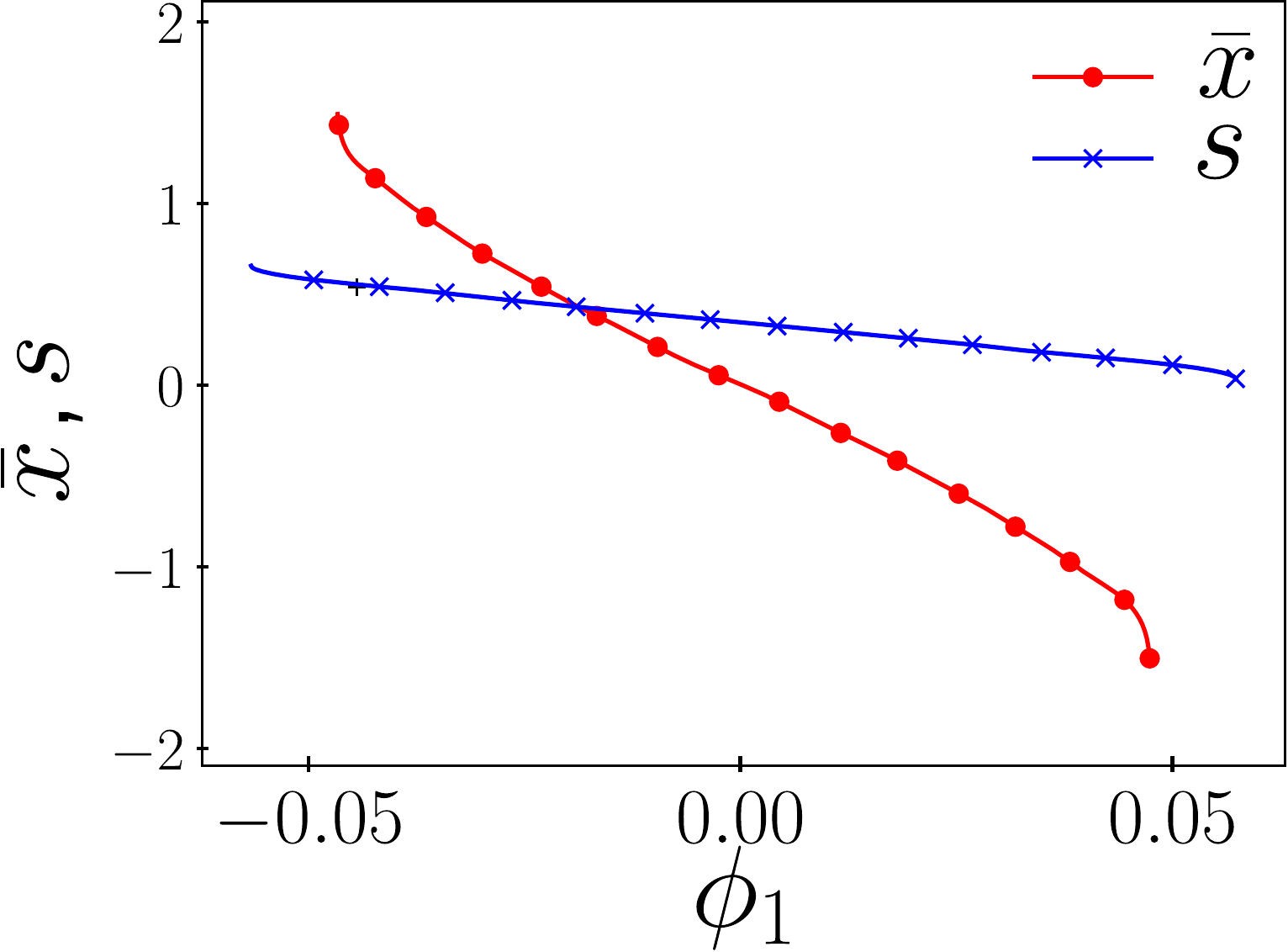} &
\includegraphics[width=0.23\textwidth]{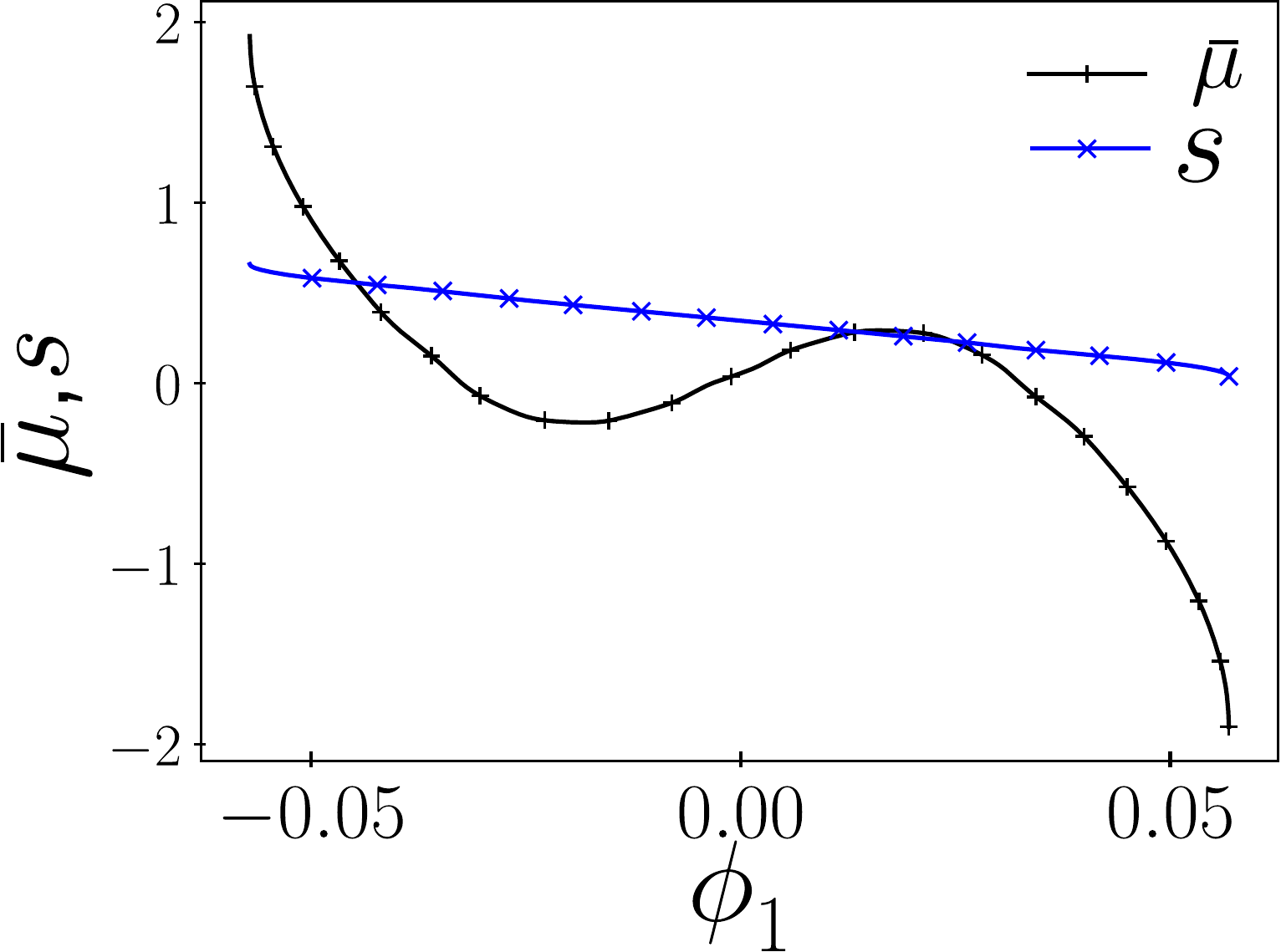} \\
(a) & (b)
\end{tabular}
\caption{Results from data mining forty trial observations (each comprising five measurements)
sampled along a single S-shaped curve. Each trial is separated from the next by
0.5\% of the total arclength.  
(a) Recording five $x$ values per segment; 
The plot shows the arclength $s$ as a function of the 
first nontrivial DMAP coordinate 
$\phi_1$, which is one-to-one with the arclength. 
The average observation value for each segment $\bar{x}$ 
is also plotted.
(b) Recording only five $\mu$ values per segment;
$\phi_1$ again provides a consistent parametrization of the curve,
even though $\mu$ is neither an injective function of $\phi_1$ nor of the arclength.}
\label{fig:x_mu_dmap}
\end{figure}

Indeed, using DMAP on the data set of vectors comprised of five consecutive $x$ observations along the blue curve in Fig.~\ref{fig:x_mu_lambda_2d} uncovers the correct (one-dimensional) topology (the right relative ordering of the trials) and provides a consistent
parametrization in terms of the first nontrivial DMAP coordinate.
Therefore, $x$ can be written as a function of the 
intrinsic variable $\phi_1$ (see Fig.~\ref{fig:x_mu_dmap}(a)).

Fig.~\ref{fig:x_mu_dmap}(b) illustrates a more challenging scenario, in which during a trial we only observe sequences of values of $\mu$ (which is {\bf not} one-to-one with the blue curve's arclength $s$). 
Still, sufficient (in the sense of Whitney) $\mu$ observations along each segment allow us to discover the intrinsic one-dimensionality,  through the parametrization via the leading diffusion map eigenvector $\phi_1$. 
Fig.~\ref{fig:x_mu_dmap}(b) confirms that $\phi_1$ {\em is} one-to-one with arclength; the S-shape of the plot of the average of the $\mu$ values in each segment, $\bar{\mu}$, as a function of $\phi_1$ provides a data-driven way to ``discover'' the system hysteresis, a key feature of the cusp surface.
One-parameter continuation can be performed ``to the left'' or ``to the right'' of a given starting point; so observing the values along a segment in reverse order is also possible.
One can take this and other symmetries into account during data mining by constructing metrics that are invariant to them (e.g. \cite{berry-2018b}).

\subsection{Reconstructing a relation}

\begin{figure}[!htp]
\centering
\includegraphics[width=0.4\textwidth]{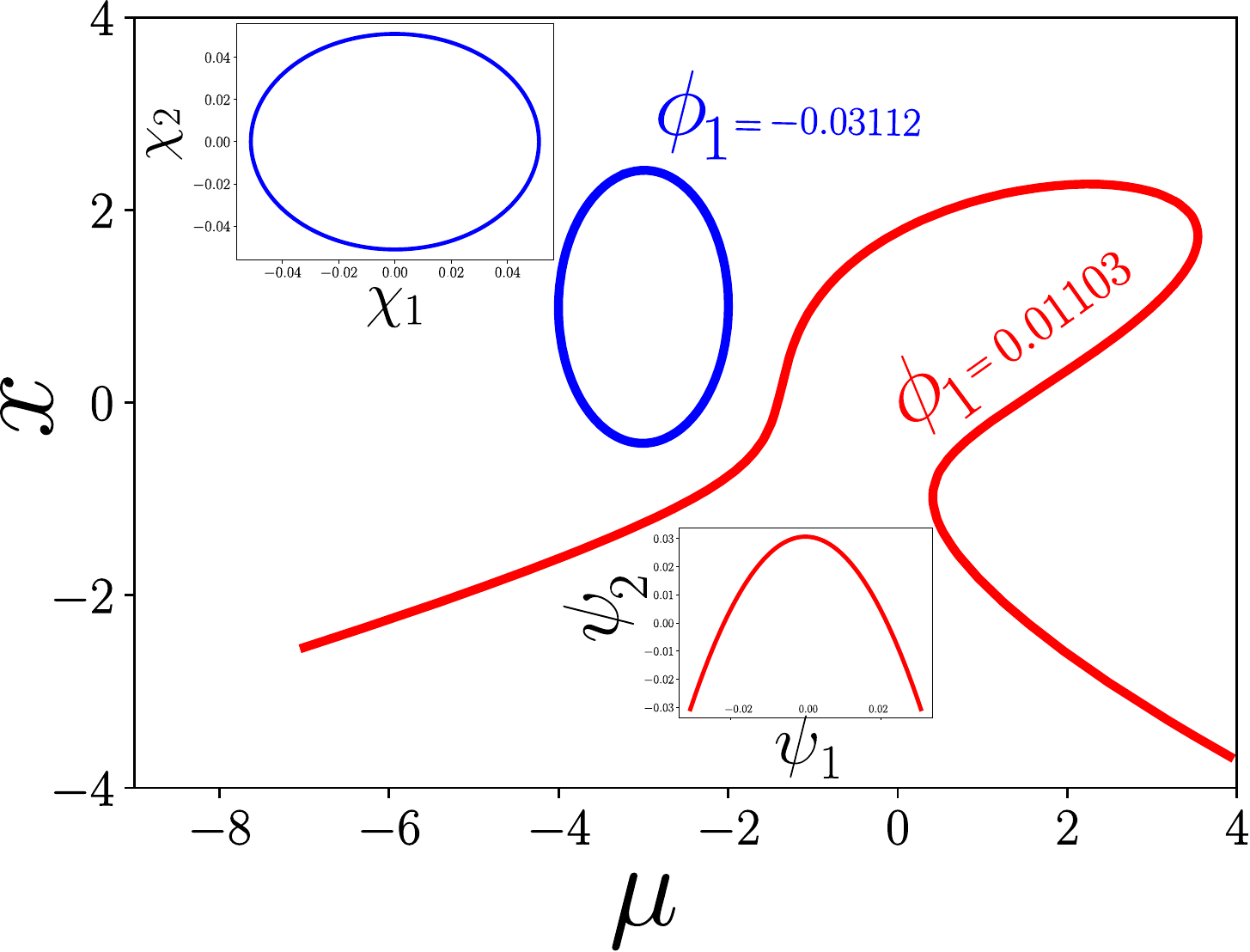}
\caption{
Multi-valued input-output (parameter-state) relations, defined through 
$0=\mu^2-2\mu x+x^3-2x-2$ and $0=(-\mu-3)^2-\exp(-0.005(2-x))+0.5(3-x)^2$.
For a single value of the parameter $\mu$ up to 
three distinct states can coexist. 
For the same state measurement up to 
four different parameter values can be found.
In this case we only record three $x$ values in each trial segment.
The original data set is colored by the first DMAP coordinate $\phi_1$.
This reveals two clusters with distinct
values of $\phi_1=0.01103,-0.03112$.
The insets show DMAPS applied on each cluster separately: 
the isola can be embedded using its first two DMAP 
eigenvectors $(\chi_1,\chi_2)$; the mushroom-like curve can 
be parameterized by its first eigenvector $\psi_1$.
The remaining DMAP coordinates (including $\psi_2$)
are harmonics of $\psi_1$.}
\label{fig:caricature_isola}
\end{figure}

Our first example was ``easy'', since an observable exists (here $x$) that is one-to-one with arclength along the overall sampled curve.
Consider now a more interesting, multivalued {\em relation} 
between input and output, such as the one shown 
in Fig.~\ref{fig:caricature_isola}, where 
the projections to {\em both} observable axes are multivalued. 
Here, we record observations of the two components (red and blue) shown in 
Fig.~\ref{fig:caricature_isola} in terms of 
one-parameter arclength segments. 
The space of observations containing sequences of only $x$ measurements 
along the curves, contains a diffeomorphic copy of the two distinct components.
Performing DMAP computations with a compactly supported kernel
reveals these two disjoint sets since
the Markov chain is reducible \cite{chung-1996,nadler-2006}.
Separately parameterizing the red and blue components reveals
the topology of each component---one is equivalent to a circle and the other to a line.
Data mining is thus capable of learning {\em relations}, 
and not just functions, discovering various disconnected components,
and then providing a useful parameterization 
of each one of them. 
We will show later in the paper that this can lead to a meaningful 
{\em transport map} between inputs and outputs that 
circumvents the complicating multivaluedness.

\subsection{Reconstructing a response surface}
\begin{figure*}
\centering
\begin{tabular}{c}
\hspace{.4cm}\includegraphics[width=0.9\textwidth]{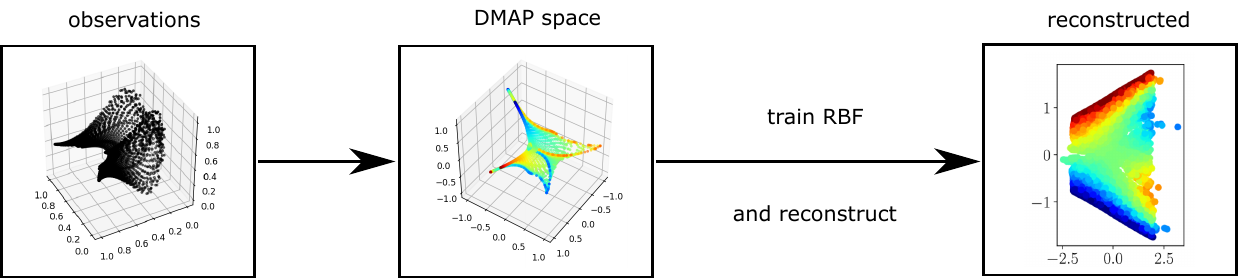}\\
(a)\\
\begin{tabular}{ccc}
\includegraphics[width=0.3\textwidth]{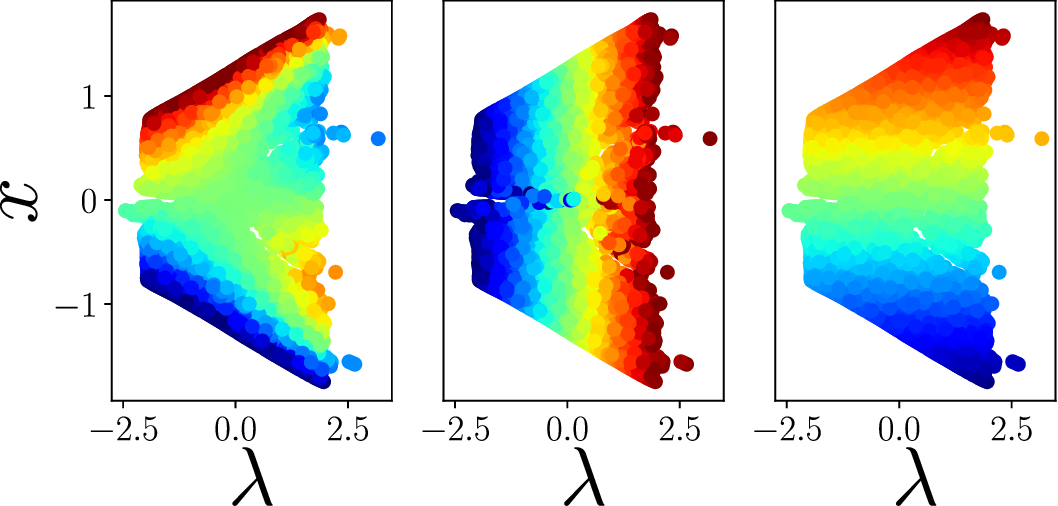}&\hspace*{1cm}&
\includegraphics[width=0.3\textwidth]{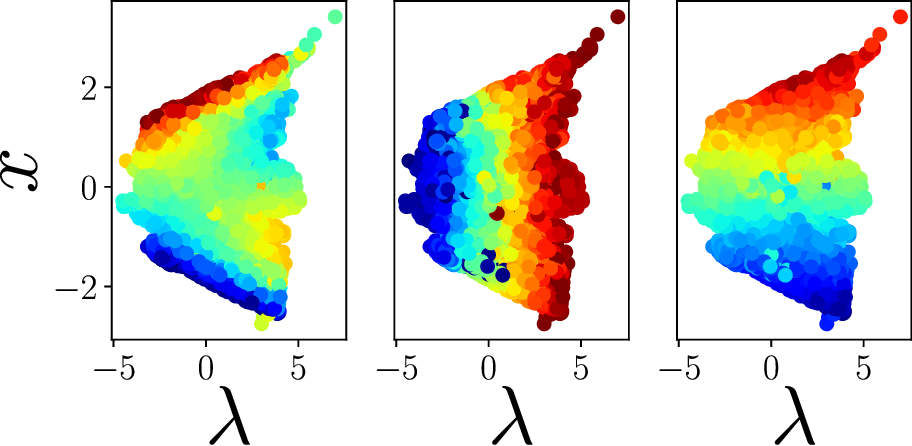}\\
(b)&\hspace*{1cm}&(c)
\end{tabular}\\
\end{tabular} 
\caption{
(a) Recording only five $x$ values from trials involving variation in the $\mu$ direction creates an embedding in observation space $\mathbb{R}^5$ (a projection of the segments to $\mathbb{R}^3$ is shown in black). Parametrizing the data with DMAPS yields a lower-dimensional embedding. We then obtain $(\mu,\lambda,x)$ as functions on the DMAP space, here by using radial basis functions with 10\% of the data used for their construction.
The plots in (b) show the reconstruction of the remaining 90\% of the data (not used for training).
(c) Reconstructions of the coordinate 
functions of the surface projected
to $(x,\lambda)$ space by recording statistics 
of 200 $\mu$ values in each of 5535 discs (see text).}
\label{fig:reconstruction_xmulam_hairs}
\end{figure*}

The third example involves the data mining of short 
one-parameter segments varying in the $\mu$ direction,
for a range of randomly distributed $\lambda$ values, sampled across
the entire surface. We only record $x$ values; $\lambda$ and $\mu$
are not explicitly measured but rather implicitly inferred.
Fig.~\ref{fig:reconstruction_xmulam_hairs} shows that 
DMAPS will recover a useful embedding of the entire two-dimensional response surface,
one that organizes the unstructured observations in the correct topology.

Picking a different bifurcation observation 
(e.g. performing short one-parameter continuations in the 
$\lambda$ direction for random initial values of $x$)
would also allow us to reconstruct the surface (not shown); 
certainly the dimension and the topology would be the same,
even though the observation process is different.

\begin{figure}
\centering
\includegraphics[width=0.4\textwidth]{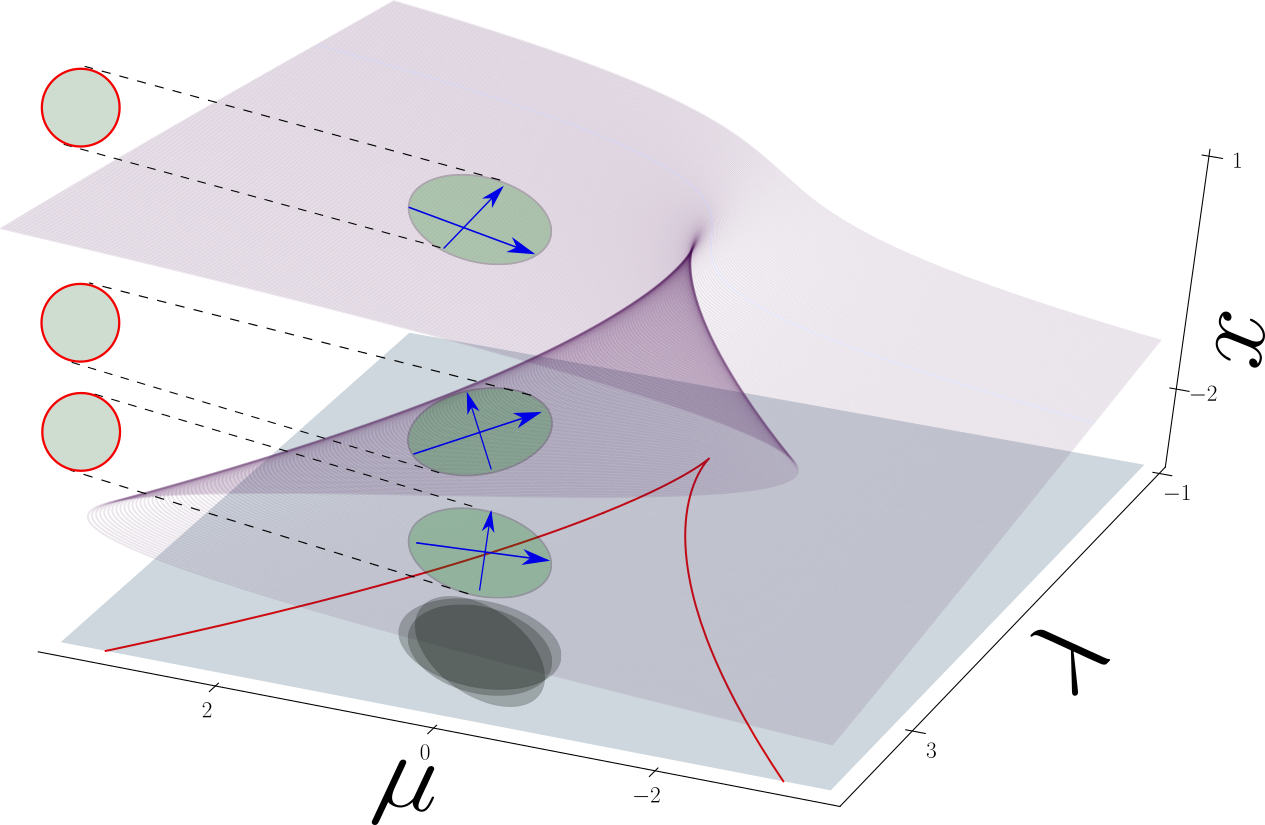} \\
\caption{Patches on the cusp surface projecting to
discs in the $(x, \lambda)$ plane,
and statistics of the $\mu$ values (denoted by arrows)
collected in these patches. The parametrization and subsequent estimation of the coordinate functions is shown in Fig.~\ref{fig:reconstruction_xmulam_hairs}(c).}
\label{fig:reconstruction_xmulam_disks}
\end{figure}

An illustration of a {\em qualitatively different} observation process,
one that is not constrained along a one-parameter path, is shown in 
Fig.~\ref{fig:reconstruction_xmulam_hairs}(c).
Now we observe the response surfaces by selecting small
patches around randomly sampled initial points (see Fig.~\ref{fig:reconstruction_xmulam_disks}).
Here, we record {\em statistics of $\mu$ values} at points uniformly 
sampled in disc-shaped patches in the $(x, \lambda)$ space.
By statistics we mean the first five moments of the distribution; other options are also possible (e.g.~Principal Component Analysis of histograms
of $\mu$ values \cite{shnitzer-2017}).
The correct dimensionality and topology, and a useful parametrization of the surface 
are again recovered, and will indeed be recovered for any sufficiently rich set of generic (alt. prevalent) 
\cite{takens1981detecting,sauer-1991} observables.
This shows that we can recover the dimension and topology, and construct a useful geometry, of the response surface even in cases where only a few quantities 
can be measured.
Different reconstructed surfaces, obtained, for example, from different types of observation of the same underlying surface, can be mapped to each other---we will return to this in the conclusions.

Takens embeddings \cite{takens1981detecting}
using temporal observations of the system state
can be formulated in either discrete or continuous time: 
one can either use values of the observable at a discrete 
number of time points, or use time derivatives of the observable
at a given time point as the additional embedding dimensions. 
The analogous observables for response surfaces
would be a state variable measurement for a single parameter setting,
and then derivatives of the observable(s) with respect to the continuation parameter, or, alternatively, with respect to the bifurcation
curve arclength at that point.
Indeed, k-jet extensions~\cite{golubitsky-1973} lift mappings to higher-dimensional spaces by replacing the value of the mapping by its Taylor series expansion of degree k. This can be viewed as an approximation of using germs or, in our case, short continuation segments to represent the map. Singularity theory characterizes when the extensions have images that are smooth manifolds.

\subsection{A chemical reactor example}
Having introduced our main idea, and demonstrated it through illustrative examples (with some of the relevant theory following in the Appendix), we now apply it in a hydrogen combustion model setting using a CSTR.


The reactor is a simplified model for the study of 
combustion dynamics, in which an extreme mixing assumption leads
to a homogeneous mixture of reactive ideal gases. 
Due to the simplicity of the model, one can investigate the contribution
of kinetics parameters on the observed dynamics.
The dependence of the reactor state, e.g.~the reactor
temperature $T$, on  the residence time in the reactor, $\tau$, typically displays 
an S-shaped curve, connecting the weakly- and strongly burning steady state 
branches via an unstable steady state branch \cite{law2006combustion}.
The turning points in the S-shaped curve correspond to ignition and extinction
limits of the mixture.
A system of $n_s$ chemical species $M_i, \,\, i=1,\dots,n_s$ reacts according
to $n_r$ reversible elementary reactions:
\begin{equation}
\sum_{i=1}^{n_{s}}\nu_{ik}^{\prime}M_{i} \rightleftharpoons 
\sum_{i=1}^{n_{s}}\nu_{ik}^{\prime\prime}M_{i},\quad k=1,\dots, n_r.
\end{equation}
Stoichiometry is defined by $\nu_{ik}^{\prime}$ and $\nu_{ik}^{\prime\prime}$, 
the stoichiometric coefficients of species $i$ in reaction $k$ for the reactants 
and products, respectively. The rate of the $k$-th elementary reaction is 
\begin{equation}
q_{k}= q^f_k - q^r_k =
k^f_k \prod_{i=1}^{n_s}[X_{i}]^{\nu_{ik}^{\prime}} 
-k^r_k \prod_{i=1}^{n_s}[X_{i}]^{\nu_{ik}^{\prime\prime}}
\qquad k=1,\dots, n_r
\end{equation}
where $[X_i]$ denotes molar concentration of species $i$
and $k^f_{k}$ and $k^r_{k}$ are the forward and reverse rate constants
of reaction $k$. The production or consumption rate $\dot{\omega}_i$ of the $i$-th species 
 is the summation of the rates of all reactions involving 
the species $i$,
\begin{equation}
\dot{\omega}_i=\sum_{k=1}^{n_r}\nu_{ik}q_k,
\end{equation}
where $ \nu_{ik}=\nu_{ik}^{\prime\prime}-\nu_{ik}^{\prime}$ is the 
net stoichiometric coefficient.

The temporal evolution of $Y_i$, the mass fraction of species $i$, and 
temperature $T$ in a perfectly stirred reactor is described by a system of $(n_s+1)$
ordinary differential equations (ODEs) \cite{kooshkbaghi2015n},
\begin{eqnarray} \label{consEqn}
\frac{dY_i}{dt} &=& \frac{1}{\tau}(Y_{i}^{0}-Y_i) + \frac{\dot{\omega}_i W_i}{\rho} 
\qquad i=1,\cdots, n_s\\ \nonumber
\frac{dT}{dt}  &=& \frac{1}{\overline{c}_p\tau} \sum_{i=1}^{n_s}(h_{i}^{0}-h_i) Y_{i}^{0} 
-\frac{1}{\rho \overline{c}_p}\sum_{i=1}^{n_s} h_i W_i\dot{\omega}_i,
\end{eqnarray}
where $Y_{i}^{0}$ and $h_{i}^{0}$ are the mass fraction and total enthalpy of 
species $i$ at the inflow,  $W_i$ and $h_i$ are the molecular weight and
total enthalpy of species $i$, and $\bar{c}_p$ and $\rho$ are the mixture heat 
capacity under constant pressure and density.
Fixing the inflow mixture composition, the reactor temperature is a function
of inlet temperature $T_0$ and residence time $\tau$.

In this study, we used a H$_2$/air ignition model in which 
the detailed kinetic mechanism includes 9 species ($n_s=9$),
participating in 21 elementary reversible reactions ($n_r=21$)
\cite{o2004comprehensive}.
AUTO-07p \cite{doedel1981auto, doedel2007auto} 
code (a bifurcation analysis tool) is coupled  with
CHEMKIN \cite{kee1996chemkin}
(a chemical kinetics database) for performing the numerical continuation computations
\cite{kooshkbaghi2015n}.

\begin{figure}
\centering
\includegraphics[width=1\columnwidth]{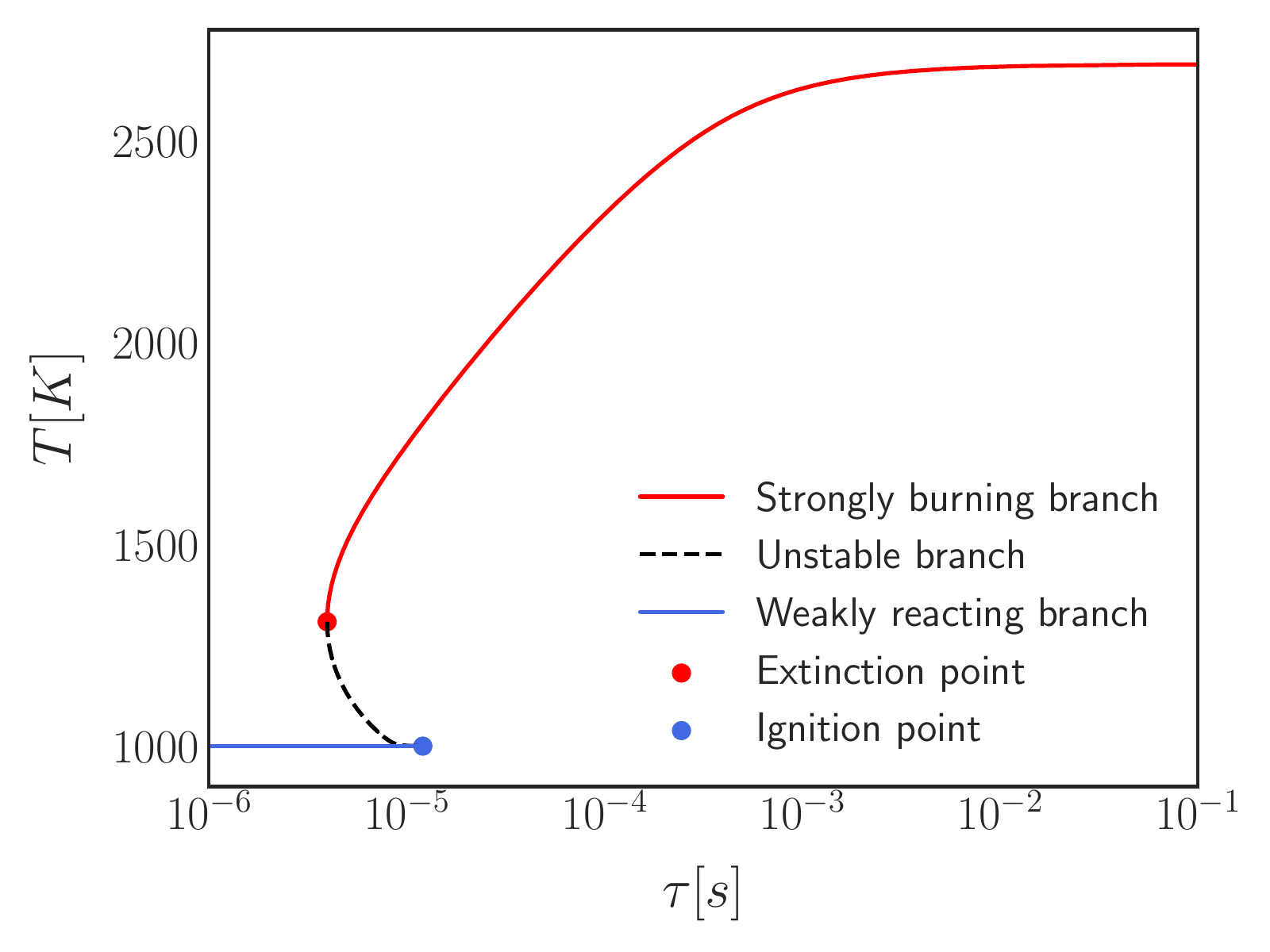}
\caption{The dependence of reactor temperature on the residence time
for stoichiometric H$_2$/air mixture with initial temperature  $T_0=1000$ K
under atmospheric pressure of one atm.
The lower branch starts from frozen state at very short
residence time and stays ``weakly reactive'' up to the ignition point. The
system state then jumps to the ``strongly burning branch''. 
If the system is already ignited, and the residence time is decreased gradually,
the system will jump back to the weakly reactive state loci at the extinction limit.
}
\label{fig:bif_T1000}
\end{figure}

The dependence of the reactor temperature $T$ on the residence time
for a stoichiometric H$_2$/air mixture 
(with pressure set to one atm, $T_0 = 1000$ K) in an adiabatic CSTR is shown in 
Fig.~\ref{fig:bif_T1000}.

\begin{figure}
\centering
\includegraphics[width=.9\columnwidth]{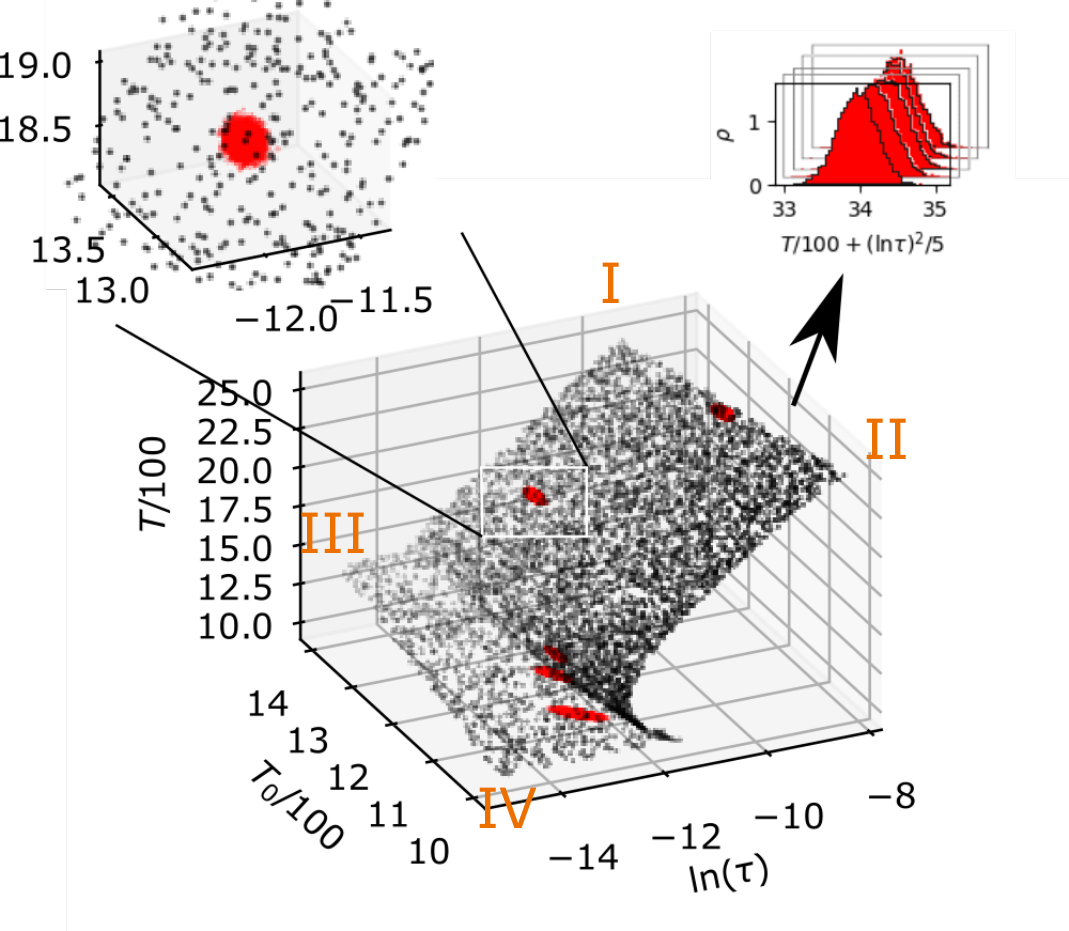}
\caption{Bifurcation surface of the CSTR, embedded in $(\ln\tau,T_0/100,T/100)$ space. The plot shows the surrounding neighborhoods (red) of five of the points on the surface, and the inset shows a zoomed-in version. The points in each red neighborhood are assembled into a histogram (top right), and we then compute the first four moments of the histogram---these four moments at each point on the surface will be the only data available for our data mining. The ``corners" of the data  set are labeled (I-IV) for easier visual reference to Fig.~(\ref{fig:cstr_embedding}).}
\label{fig:cstr_sampling}
\end{figure}

\begin{figure}
\centering
\includegraphics[width=1\columnwidth]{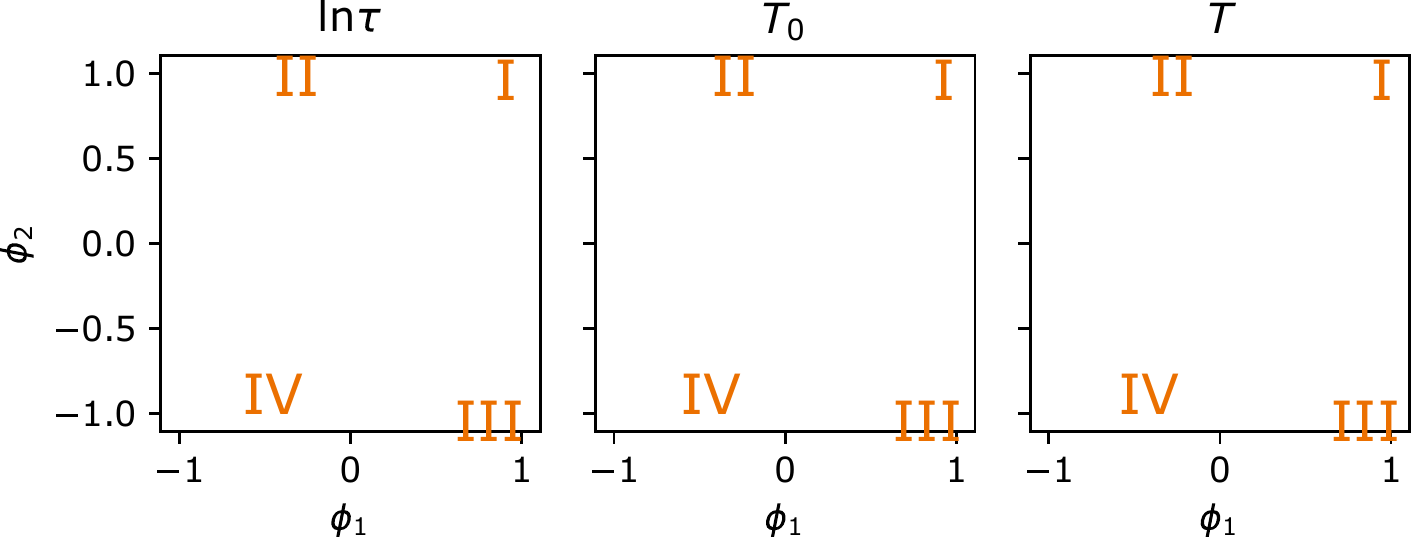}
\caption{Embedding of the moment data in Diffusion Map space, colored by the log residence time $\ln\tau$, initial temperature $T_0$, and steady state temperature $T$. The two clusters in the point cloud belong to the regions before and after the fold on the diagram. The labels of the edges (I-IV) correspond to the labels in Fig.~(\ref{fig:cstr_sampling}).}
\label{fig:cstr_embedding}
\end{figure}

We assume that an experimenter is able to initialize the reactor at many different values of the initial temperature $T_0$ and residence time $\ln\tau$, but without necessarily having knowledge of the numerical value of these parameters. 
We also assume that the initialization at a fixed parameter value of $(T_0,\ln\tau)$ can be repeated, each time with a normally distributed inaccuracy $\delta$ in the position on the surface (see Fig~\ref{fig:cstr_sampling}). 
After initialization, the experimenter then records real-valued observations. 
For the computational experiment, we rather arbitrarily choose $y=(\ln\tau)^2/5+T$, a combination of $\tau$ and the temperature $T$ at steady state as the observed/recorded quantity.  
Notably, the initial temperature $T_0$ is not part of the observation---in fact, by theorem~\ref{thm:whitney prevalence} (Appendix A), almost any one--dimensional combination of the variables is admissible.
We initialize $5535$ points on the surface with rescaled coordinates $(\ln\tau,T_0/100, T/100)\in [-15,-8]\times [10,14]\times [10,24]$.
Similar to the illustration in Fig.~\ref{fig:reconstruction_xmulam_disks}, at any given point $p$ out of the 5535 points on the surface, we record values of $y=(\ln\tau)^2/5+T$ for 10000 points $(p+\delta)$, i.e. in a small neighborhood of $p$ on the surface (red ellipses in Fig.~\ref{fig:cstr_sampling}). 
The first four moments (the mean, standard deviation, skew, and kurtosis) of the values of $y$ are used as the measured features of the trial associated with the point $p$. 
This leads to a dataset $X\in\mathbb{R}^{5535\times 4}$. We then apply Diffusion Maps with kernel bandwidth $\epsilon=1.25$ (see refs.~\cite{berry-2016b,singer-2006x} for a discussion of the choice of kernel bandwidth) to parametrize the manifold embedded in this space. Fig.~\ref{fig:cstr_embedding} shows empirical evidence that a diffeomorphic copy of the two-dimensional response surface for this reactor can be recovered, even though we only had access to a few moments of the scalar observations $y$ on the surface
(a physically rather unusual observation choice!).

\section{Vector fields and transport maps}
\begin{figure}
\centering
\begin{tabular}{cc}
\includegraphics[width=0.22\textwidth]{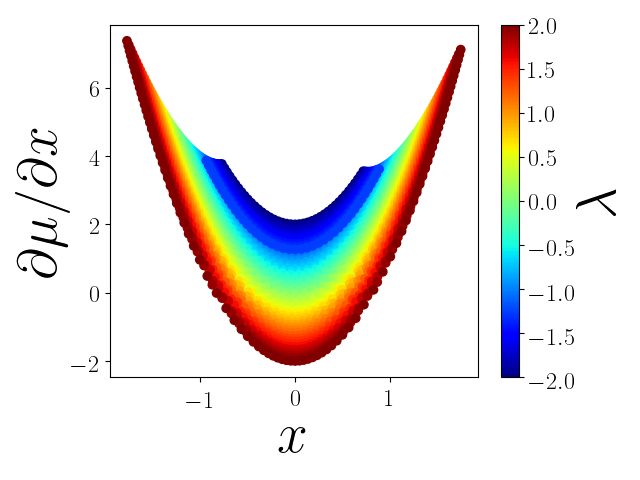} &
\includegraphics[width=0.22\textwidth]{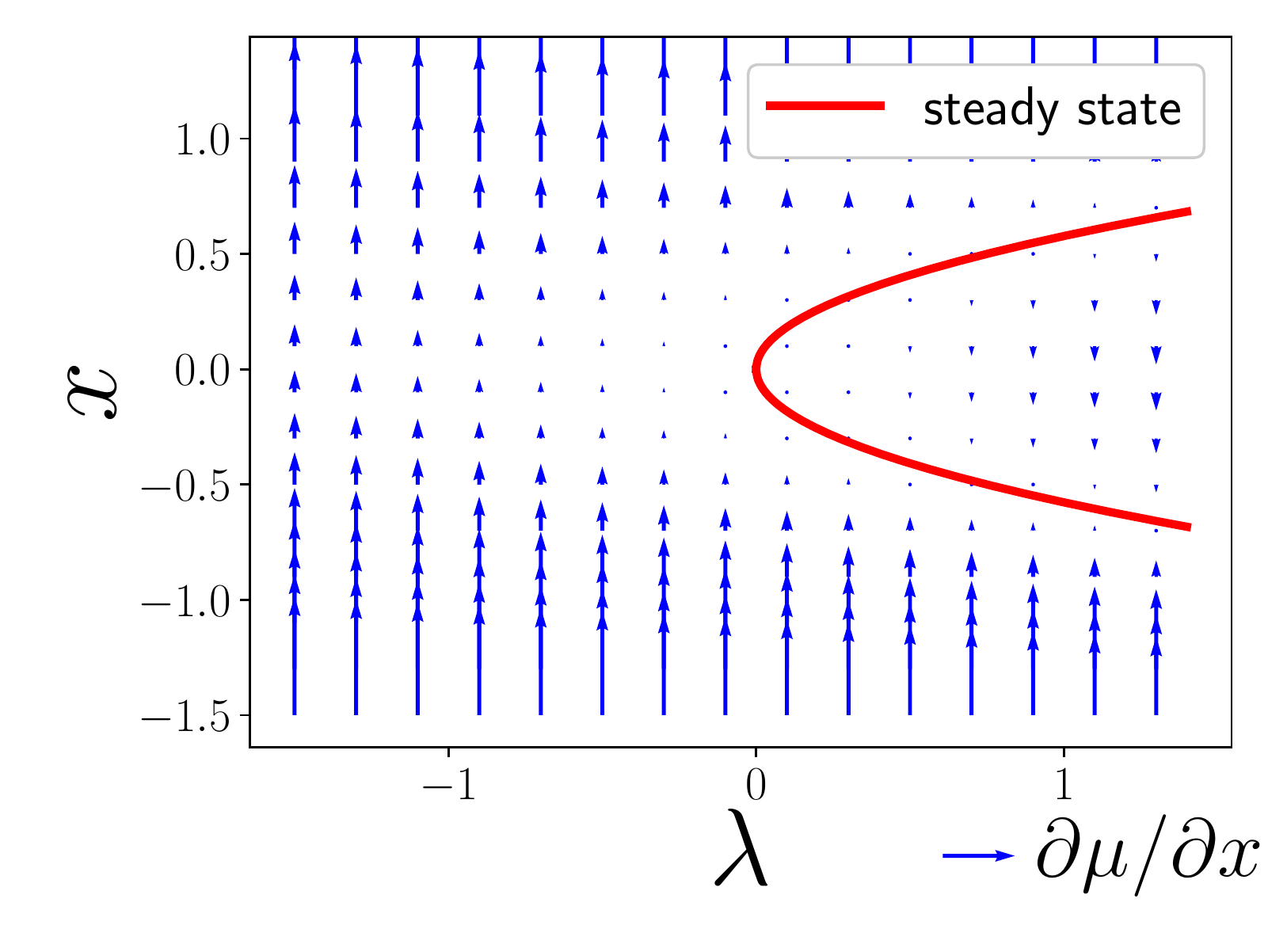} \\
(a) & (b)
\end{tabular}
\caption{(a) Embedding of the cusp surface in 
$x$, $\partial \mu/\partial x$ for each fixed $\lambda$.
$\lambda$ can be written as a function over this 
embedding (color).
This shows that we can recover 
the second parameter without measuring
it directly and just by local information 
attained by change in the first parameter. 
(b) A one-dimensional vector field 
over $x$, parametrized by $\lambda$.
The vectors $\partial \mu/ \partial x$
at each $x$ (the arrows) are given by $3x^2-\lambda$.
Two steady state branches on which the vector field is zero are ``born'' at the 
turning point, corresponding to the original cusp point.}
\label{fig:vectorfield}
\end{figure}

An embedding of the cusp surface in two-dimensional
$(x, \partial \mu/\partial x)$-space is shown in 
Fig.~\ref{fig:vectorfield}.
It is maybe interesting that this response surface embedding 
can be alternatively interpreted 
as an observation of a one-dimensional vector field 
$\mathbf{V}$ over $x$,
parametrized by $\lambda$, 
i.e.~$\mathbf{V}:= \frac{\partial\mu}{\partial x}(x;\lambda)$; or, equivalently, as the 
right-hand-side of an ordinary differential equation,
$\frac{d}{dt}x(t) = V(x(t);\lambda)$.
As shown in Fig.~\ref{fig:vectorfield}(b) the turning 
points of the cusp surface become steady 
state branches for the dynamics of this vector field. 
The cusp point now describes a saddle-node bifurcation.
Short continuations in the $\mu$ direction
(for fixed $\lambda$) correspond to {\em time-series segments} for this constructed vector field.
``Time'' for this vector field goes to infinity
at finite $\mu$ (at the turning points, which now become steady states).

Clearly, tools for signal-processing and
identification of dynamical systems 
(e.g.~Koopman operator approximations \cite{budisic-2012,williams-2015,bollt-2018b},
time-delay embedding techniques, 
information theory \cite{bandt-2002,talmon-2013}, etc.)
can be exploited to develop useful observables for continuation-based bifurcation surface reconstruction.

\begin{figure*}[!htp]
\centering
\begin{tabular}{cc}
\includegraphics[width=0.3\textwidth]{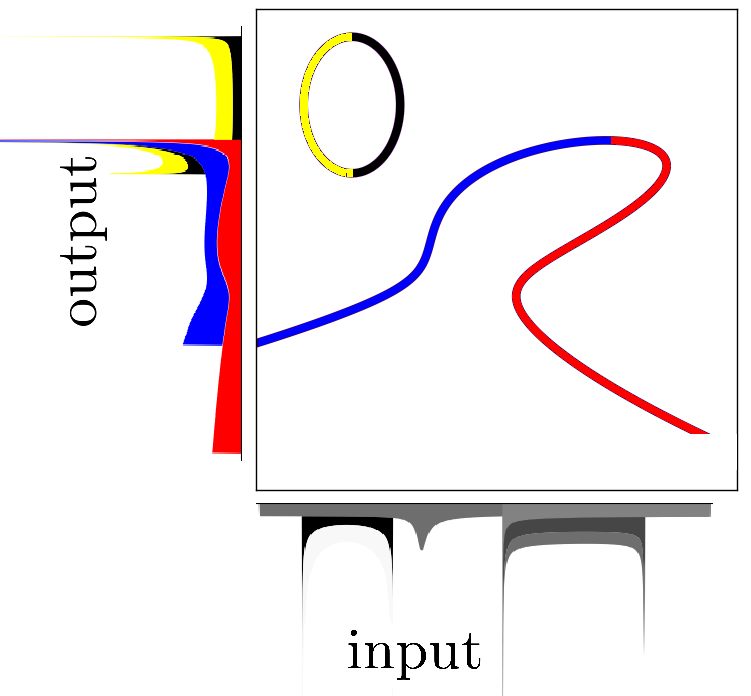}&
\includegraphics[width=0.5\textwidth]{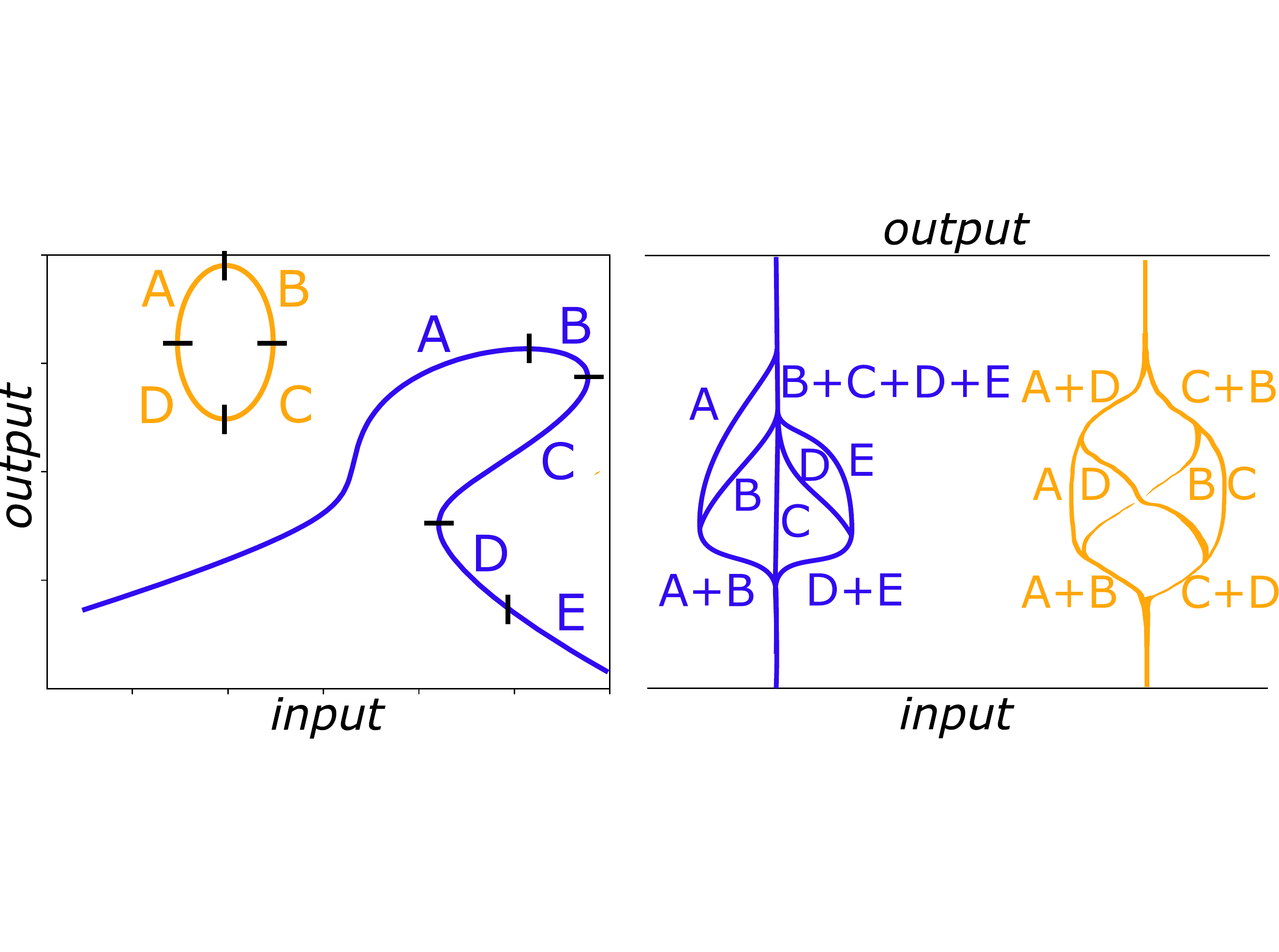}\\
(a)&(b)
\end{tabular}
\caption{\label{fig:branching isola}(a) Uniform density on the arclength of each of two curves, 
transported through the relation to the two axes. 
The colors (yellow and black for the isola, red and blue for the ``mushroom'') 
show which part of the resulting density is caused by 
which part of the relation. The black density on the horizontal axis shows an equivalent transport (with the corresponding colored parts on the curve not indicated here).
(b) One-to-one sub-branches of the relation, with branching during the transport. The colors on the right panel follow from the colors of the curves in the center panel. The curves are split into more segments than the four colored ones on the left panel, because the branched transport can be considered in both the input-output and the output-input directions.}
\end{figure*}

The process of observing response surfaces also 
has interesting implications in studying transport maps 
between inputs and outputs~\cite{evans-1999}.
Consider the relation we explored 
in Fig.~\ref{fig:caricature_isola}, 
and consider an observation process that samples the 
arclength of each of its two component curves at equal steps; this 
produces a constant density of observations 
along the arclength of each component.
Projecting this uniform density onto the input and the
output axes results in complicated density profiles, 
with the density approaching infinity at turning points of the curve
(spikes in Fig.~\ref{fig:branching isola}).
Detection of the location of such singularities in a higher-dimensional
setting is also possible through data-driven analysis of the graph Laplacian on the data~\cite{belkin-2012}.
Given the densities on input and output axes,
it is natural to explore a transport map between them. 
Instead of trying to construct such a transport map through an 
appropriately defined optimization process
\cite{wasserstein-1969,villani-2009},
we can use data mining of the bifurcation observations themselves
to provide a meaningful solution.
In the example shown in Fig.~\ref{fig:branching isola}, 
DMAP can uncover and parametrize the two 
disconnected components, allowing us to write the 
relation between input and output {\em in a parametric form in terms of DMAP coordinates}.
Since the relation embodies the transport map, this
helps us effectively recover it in the same parametric form. 
Projecting the density on the intrinsic parametrization separately 
to the input axis and then to the output axis demonstrates how the 
singularities in input and output densities form: the projection of the uniform density $\rho_s$ along the arclength $s$, results in the density $\rho_y$ on the output (vertical, $y$) axis. The result $\rho_y$ is seen, in this case, to be the sum of transports over piecewise invertible branch segments $g_k:\mathbb{R}\to\mathbb{R}$, $k=1,\dots,4$ of the curves (differentiated by color in Fig.~\ref{fig:branching isola}):
\begin{equation}
\rho_y(y):=\sum_{k=1}^4 \left|\frac{\partial g_k}{\partial s}(g_k^{-1}(y))\right|^{-1}\rho_s(g_k^{-1}(y)).
\end{equation}
This formula encodes the action of a type of transfer operator on the density $\rho_s$ (see, e.g. \cite{baladi-2000,ruelle-2012,bollt-2013}).

Similar considerations apply to the projection on the ``input'' $x$ axis.
The book-keeping introduced through the intrinsic parametrization
allows us to determine how the input and output densities 
are connected through a form of branched transport (see Fig.~\ref{fig:branching isola}).
%
Gilbert et. al~\cite{gilbert-1967, gilbert-1968} discussed branched transport in the solution to Steiner's minimal path problem with atomic measures, and Xia~\cite{xia-2003} generalized the concept to arbitrary probability measures.
It is important to state that this parametrization of 
the transport map has required more than just
single point observations. 
The additional information necessary to embed and parametrize 
the ``relation manifold'' is gained through {\em the observation process}, i.e. the context provided in each set of observations.

\section{Conclusions}
In summary, we have demonstrated that observations of input-output relations,
in the form of ensembles of disorganized, even partially recorded,
short ``trials'' 
can be rearranged and combined to construct a realization 
(and parametrization) of the full response
surface through manifold learning techniques. 
This relies crucially on the Whitney and Takens embedding theorems,
typically used in time series analysis of dynamical systems, 
now applied to one-parameter continuation for 
the construction of response surfaces.
We also demonstrated that 
different types of observations (not just single-parameter 
continuation segments) can be used similarly.
It is possible (though it is not shown here), 
to compute Diffusion Maps not with the Euclidean distance 
between observations, but with a Mahalanobis-like metric
that takes into account local observation covariances
\cite{singer-2008,dsilva-2016}.
This has the potential of fusing different observation sets
(observations from one-parameter continuation runs in
different directions, statistics from two-dimensional patches,
etc.) in a single ``master'' observation surface, but requires a consistent way to estimate
covariance matrices  for the given observation process.
Our approach can then be useful in a domain-adaptation context, by constructing meaningful realizations of the input and output (``source'' and ``target'') domains.
It would be interesting to compare the Mahalanobis-like metric driven observation fusion with other registration approaches developed in the domain-adaptation literature~\cite{courty-2017,yair-2019}. 

There is a conceptual similarity to the Dynamic Laplacian~\cite{Froyland-2017}, in that, to gather covariance information, we have to start at several nearby initial trial points 
and then perform the trial associated with each one of them.
In addition, we explored how the data driven recovery of the
dimensionality and topology/geometry of the input-output relation 
can lead to a useful transport map 
between parameter (input) and state (output) spaces.
Current work explores extending the approach to other response surface types, as well as the relation of this ``observation based'' transport map to those obtained through optimal transport considerations.

\begin{acknowledgments}
The work was partially funded by the National Science Foundation, DARPA (IGK, FD), SNSF grant P2EZP2\_168833 (MK), and ARO and ONR (EB). Discussions with Prof.~J.~Guckenheimer are gratefully acknowledged.
\end{acknowledgments}

\bibliographystyle{apsrev4-1}
\bibliography{bif_recover}

\appendix
\section{Appendix A}
Let $k\geq d\in\mathbb{N}$, and consider a manifold $\mathcal{M}\subset\mathbb{R}^k$ that is $d$-dimensional, compact, smooth, connected, oriented and endowed with a Riemannian metric $g$ induced by the embedding in $k$-dimensional Euclidean space.
Note that this setting is sufficient for our presentation, but is more restrictive than allowed by the results cited below.
Together with the results from Packard et al.~\cite{packard1980geometry} and Aeyels~\cite{aeyels-1981}, the definitions and theorems of Takens~\cite{takens1981detecting} describe the concept of observability of state spaces of nonlinear dynamical systems. A dynamical system is defined through its state space (here, the manifold $\mathcal{M}$) and a diffeomorphism $\phi:\mathcal{M}\to\mathcal{M}$.
\begin{theorem}\textbf{Generic delay embeddings.}\label{thm:takens 1}
For pairs $(\phi,y)$, $\phi:\mathcal{M}\to\mathcal{M}$ a smooth diffeomorphism and $y:\mathcal{M}\to\mathbb{R}$ a smooth function, it is a generic property that the map $\Phi_{(\phi,y)}:\mathcal{M}\to\mathbb{R}^{2d+1}$, defined by
\begin{equation}
\Phi_{(\phi,y)}(x)=\left(y(x),y(\phi(x)),\dots,y(\underbrace{\phi\circ\dots\circ\phi}_{2d~\text{times}}(x))\right)
\end{equation}
is an embedding of $\mathcal{M}$; here, ``smooth'' means at least $C^2$.
\end{theorem}
Genericity in this context is defined as ``an open and dense set of pairs $(\phi,y)$'' in the $C^2$ function space.
In Takens' paper, there is also an infinitesimal version of theorem~\ref{thm:takens 1}:
\begin{theorem}\textbf{Generic differential embeddings.}\label{thm:takens 2}
For pairs $(X,y)$, $X:\mathcal{M}\to T\mathcal{M}$ a smooth vector field with flow $\phi_t:\mathcal{M}\to\mathcal{M}$, and $y:\mathcal{M}\to\mathbb{R}$ a smooth function, it is a generic property that the map $\Phi_{(X,y)}:\mathcal{M}\to\mathbb{R}^{2d+1}$, defined by
\begin{equation}
\Phi_{(X,y)}(x)=\left(y(x),\left.\frac{d}{dt}y(\phi_t(x))\right|_{t=0},\dots,\left.\frac{d^{2d}}{dt^{2d}}y(\phi_t(x))\right|_{t=0}\right)
\end{equation}
is an embedding of $\mathcal{M}$; here, ``smooth'' means at least $C^{2d+1}$.
\end{theorem}
For an extension of Takens' theorems to deterministically forced, input-output, irregularly sampled, and stochastic systems, we refer the reader to the results from Stark et al.~\cite{stark-1997,stark-1999,stark-2003}.

Generic (open and dense) sets can have measure zero, so Sauer et al.~\cite{sauer-1991} refined the results significantly by introducing the concept of prevalence (a ``probability one'' analog in infinite dimensional spaces).
\begin{definition}
A Borel subset $S$ of a normed linear space $V$ is \textbf{prevalent} if there is a finite-dimensional subspace $E$ of $V$ such that for each $v\in V$, $v + e$ belongs to $S$ for (Lebesgue-) almost every $e$ in $E$.
\end{definition}
Using this notion one can strengthen the result from the original theorem of Whitney~\cite{whitney-1936}, into the prevalence form:
\begin{theorem}\textbf{Whitney (weak form).}\label{thm:whitney weak}
The set $S\subset C^1$ of smooth maps $F:\mathbb{R}^k\to\mathbb{R}^{2d+1}$ that are embeddings of $\mathcal{M}$ is an open and dense set in the $C^1$-topology.
\end{theorem}
\begin{theorem}\textbf{Whitney (with prevalence).}\label{thm:whitney prevalence}
The set $S\subset C^1$ of smooth maps $F:\mathbb{R}^k\to\mathbb{R}^{2d+1}$ that are embeddings of $\mathcal{M}$ is prevalent.
\end{theorem}
In particular, given any smooth map $F$, not only are there maps arbitrarily near F that are embeddings (which is the notion of genericity from Takens), but ``almost all'' (in the sense of prevalence) of the maps near $F$ are embeddings.
The space $E$ from the definition of prevalence in Thm.~\ref{thm:whitney prevalence} is the $k(2d + 1)$-dimensional space of linear maps from $\mathbb{R}^k$ to $\mathbb{R}^{2d+ 1}$.

For completeness, we also add the statement of the following, strong form of the Whitney theorem.
\begin{theorem}\textbf{Whitney (strong form, existence).}\label{thm:whitney strong}
For every $d$-dimensional manifold $\mathcal{M}$ of the form given above there exists an embedding into $\mathbb{R}^{2d}$.
\end{theorem}
Note that Thm.~\ref{thm:whitney strong} is a stronger statement than Thm.~\ref{thm:whitney weak} in terms of the  dimension of the embedding space: ($\mathbb{R}^{2d}$ instead of $\mathbb{R}^{2d+1}$); yet is not as relevant in practice, since it does not have the same probabilistic notion of prevalence.

\end{document}